\documentclass[12pt,a4paper,dvips,epsfig,openany]{article}
\usepackage{a4p,rotating}
\usepackage{epsfig,subfigure}
\usepackage{cite,mcite}
\usepackage{graphicx}
\usepackage{physics,fleqn}
\usepackage{l3_title,ifthen,Lep}
\journalname{Phys. Lett. B}
\date{September 27, 1999}
\preprint{99-131}
\newlength{\capindent}
\newlength{\capwidth}
\newlength{\figwidth}
\newcommand{\icaption}[2][!*!,!]{\hspace*{\capindent}%
  \begin{minipage}{\capwidth}
    \ifthenelse{\equal{#1}{!*!,!}}%
      {\caption{#2}}%
      {\caption[#1]{#2}}
  \end{minipage}}
\newcommand{\amp}[2]{\ensuremath{_{-#1}^{+#2}}}

\renewcommand{\eV}{\mathrm{e\kern -0.1em V}}
\renewcommand{\MeV}{\mathrm{Me\kern -0.1em V}}
\renewcommand{\GeV}{\mathrm{Ge\kern -0.1em V}}
\renewcommand{\TeV}{\mathrm{Te\kern -0.1em V}}

\renewcommand{\d}{\ensuremath{\mathrm{d}}}
\newcommand{\SIG}{\ensuremath{\mathrm{sig}}}
\newcommand{\BG}{\ensuremath{\mathrm{bg}}}

\newcommand{\Dk}{\ensuremath{\Delta\kappa}}
\newcommand{\Dkg}{\ensuremath{\Dk_\gamma}}
\newcommand{\DkZ}{\ensuremath{\Dk_{\mathrm{Z}}}}
\newcommand{\kg}{\ensuremath{\kappa_\gamma}}
\newcommand{\kZ}{\ensuremath{\kappa_{\mathrm{Z}}}}
\newcommand{\Lg}{\ensuremath{\lambda_\gamma}}
\newcommand{\LZ}{\ensuremath{\lambda_{\mathrm{Z}}}}
\newcommand{\giZ}{\ensuremath{g_1^{\mathrm{Z}}}}
\newcommand{\gvZ}{\ensuremath{g_5^{\mathrm{Z}}}}
\newcommand{\aWf}{\ensuremath{\alpha_{\mathrm{W}\Phi}}}
\newcommand{\aBf}{\ensuremath{\alpha_{\mathrm{B}\Phi}}}
\newcommand{\aW}{\ensuremath{\alpha_{\mathrm{W}}}}

\newcommand{\fit}{\ensuremath{{\mathrm{fit}}}}
\newcommand{\gen}{\ensuremath{{\mathrm{gen}}}}

\newcommand{\pz}{\ensuremath{\phantom{0}}}
\newcommand{\pzz}{\ensuremath{\phantom{00}}}
\renewcommand{\MZ}{\ensuremath{M_{\mathrm{Z}}}}
\renewcommand{\MW}{\ensuremath{M_{\mathrm{W}}}}

\newcommand{\PW}{\ensuremath{\mathrm{W}}}
\newcommand{\WW}{\ensuremath{\mathrm{WW}}}

\newcommand{\EE}{\ensuremath{\mathrm{e}^+\mathrm{e}^-}}

\newcommand{\NN}{\ensuremath{\nu\bar\nu}}

\newcommand{\EN}{\ensuremath{\mathrm{e}\nu}}

\newcommand{\LN}{\ensuremath{\ell\nu}}

\newcommand{\EENN}{\EE\ensuremath{\rightarrow}\NN}

\newcommand{\QQEN}{\ensuremath{qq\mathrm{e\nu}}}
\newcommand{\QQMN}{\ensuremath{qq\mu\nu}}
\newcommand{\QQTN}{\ensuremath{qq\tau\nu}}
\newcommand{\QQLN}{\ensuremath{qq\ell\nu}}
\newcommand{\LNLN}{\ensuremath{\ell\nu\ell\nu}}

\newcommand{\QQQQ}{\ensuremath{qqqq}}

\newcommand{\EEWW}{\EE\ensuremath{\rightarrow}\WW}

\def\p1mu{\mathrm{p^{\mu}_{1}}}

\begin{document}
\begin{titlepage}
  \title{{\Huge
      Measurement of Triple-Gauge-Boson \\
      Couplings of the W Boson at LEP \\}}

\author{The L3 Collaboration}

%
%
\begin{abstract}
  
  We report on measurements of the triple-gauge-boson couplings of the
  W boson in $\EE$ collisions with the L3 detector at LEP.  W-pair,
  single-W and single-photon events are analysed in a data sample
  corresponding to a total luminosity of 76.7~pb$^{-1}$ collected at
  centre-of-mass energies between $161~\GeV$ and $183~\GeV$.
  CP-conserving as well as both C- and P-conserving triple-gauge-boson
  couplings are determined.  The results, in good agreement with the
  Standard-Model expectations, confirm the existence of the self
  coupling among the electroweak gauge bosons and constrain its
  structure.

\end{abstract}
%
%
\submitted
\end{titlepage}
%
%
\section{Introduction}
\label{sec:intro}

During the 1996 and 1997 data taking periods, the centre-of-mass
energy, $\sqrt{s}$, of the $\EE$ collider LEP at CERN was increased
from $161~\GeV$ to $172~\GeV$ and $183~\GeV$.  These energies are well
above the kinematic threshold of W-boson pair production,
$\EE\rightarrow\mathrm{W^+W^-}$.

W-pair production, single-W production ($\EE\rightarrow\PW\EN$) 
and single-photon production ($\EENN\gamma$) all depend on the
trilinear self couplings among the electroweak gauge bosons $\gamma$,
W and Z~\cite{LEP2YRSM}.  The non-Abelian gauge structure of the
electroweak theory implies the existence of the triple-gauge-boson
vertices $\gamma$WW and ZWW~\cite{standard_model}.

To lowest order within the Standard Model~\cite{standard_model} (SM),
three Feynman diagrams contribute to W-pair production, the
$s$-channel $\gamma$ and Z-boson exchange and the $t$-channel
$\nu_{\e}$ exchange.
The $s$-channel diagrams contain the $\gamma$WW and ZWW vertices.  At
present centre-of-mass energies, single-W production is sensitive to
the electromagnetic gauge couplings only.  The $\gamma$WW vertex
appears in one of the contributing $t$-channel Feynman diagrams in
$\PW\EN$ production, and dominates the corresponding diagram
containing the ZWW vertex.  Radiation of a photon from the $t$-channel
exchanged W boson in the process
$\EE\rightarrow\mathrm{\nu_e\bar\nu_e}$ becomes significant for
centre-of-mass energies far above the Z pole and involves as well the
$\gamma$WW vertex.

In general the vertices $\gamma$WW and ZWW are parametrised in terms
of seven triple-gauge-boson couplings (TGCs) each~\cite{Hagiwara87},
too many to be measured simultaneously.  Regarding only CP-conserving
couplings and assuming electromagnetic gauge invariance, six TGCs
remain, which are $\giZ$, $\gvZ$, $\kg$, $\Lg$, $\kZ$ and $\LZ$.
Within the SM, $\giZ=\kg=\kZ=1$ and $\gvZ=\Lg=\LZ=0$ at tree level.
Except $\gvZ$ these TGCs also conserve C and P separately.

Assuming custodial $\mathrm{SU(2)}$ symmetry leads to the constraints
$\DkZ = \Delta\giZ - \Dkg \tan^2\theta_w$ and $\LZ =
\Lg$~\cite{DKL,DXY,Schildknecht96,LEP2YRAC}, where $\Delta$ denotes
the deviation of the TGC from its SM value and $\theta_w$ is the weak
mixing angle.  When these constraints are applied, the remaining three
TGCs, $\giZ$, $\kg$ and $\Lg$, correspond to the operators in a linear
realization of a gauge-invariant effective Lagrangian that do not
affect the gauge-boson propagators at tree level~\cite{LEP2YRAC}.  The
TGCs are related to the three $\alpha$ couplings used in our previous
publications~\cite{l3-111,l3-130}.\footnote{The relations
  are~\cite{LEP2YRAC}: $\aWf = \Delta\giZ \cos^2\theta_w$, $\aBf =
  \Dkg - \Delta\giZ\cos^2 \theta_w$, and $\aW = \Lg$.}

Alternatively, it is interesting to study the TGCs $\giZ$, $\kg$ and
$\kZ$, imposing $\Lg=\LZ=0$.  This set corresponds to the operators of
lowest dimensionality in the non-linear realization of a
gauge-invariant effective Lagrangian, necessary in the absence of a
light Higgs boson~\cite{LEP2YRAC}.

In this article we report on measurements of TGCs of the W boson in
data samples corresponding to total luminosities of 10.9~pb$^{-1}$,
10.3~pb$^{-1}$ and 55.5~pb$^{-1}$ collected at centre-of-mass energies
of $161~\GeV$, $172~\GeV$ and $183~\GeV$, respectively.  The results
on TGCs are based on analyses of multi-differential cross sections in
W-pair, single-W and single-photon production.  They include and
supersede our previously published results on
TGCs~\cite{l3-111,l3-130,l3-112+150}.  Other experiments at LEP and at
hadron colliders have also reported results on
TGCs~\cite{alephtgc,delphitgc,opaltgc,ACPBARP}.

\section{Event Selection and Reconstruction}
\label{sec:events}

The event selections used here are identical to those published
earlier on W-pair production~\cite{l3-111,l3-120,l3-155}, single-W
production~\cite{l3-112+150}, and single-photon
production~\cite{l3-124+160}.  The same signal and background Monte
Carlo and detector simulations are used.  The number of selected
events and the expected background are reported in
Table~\ref{tab:events-all}.

\subsection{W-Pair Events}
\label{sec:WW}

Each W boson decays into a fermion and an antifermion, for short
denoted as $qq$ or $\ell\nu$ in the following.  The visible particles
in the final state, \ie, electrons, muons, $\tau$ jets corresponding
to the visible $\tau$ decay products, and hadronic jets corresponding
to quarks are reconstructed~\cite{l3-111,l3-120,l3-155}.  For $\QQQQ$
and $\QQLN$ events, energy-momentum conservation and equal mass for
the two W bosons are used as constraints in a kinematic fit to
determine the kinematics of all four final-state fermions with
improved resolution~\cite{l3-171}.

In the case of $\QQQQ$ events, a combinatorial ambiguity arises in the
assignment of jets to W bosons.  The four jets are paired to two W
bosons following the criterion of smallest mass difference between the
W candidates, excluding the combination with the smallest sum of W
masses.  On Monte Carlo events, the resulting pairing is found to be
correct for 74\% of all selected $\QQQQ$ events at $\sqrt{s} =
183~\GeV$~\cite{l3-155}.

Summing over final-state fermion helicities, fixing the mass of the W
boson and neglecting photon radiation, five phase-space angles
completely describe the four-fermion final state from W-pair decay for
unpolarised initial states.  These are the polar scattering angle of
the W$^-$ boson, $\Theta_{\PW}$, and the polar and azimuthal decay
angles in the rest systems of the two decaying W bosons, $\theta_\pm$
and $\phi_\pm$, for the fermion in W$^-$ and the antifermion in W$^+$
decay.  TGCs affect the total production cross section, the
distribution of the W-boson polar scattering angle, and the
polarisations of the two W bosons, analysed in the distributions of
the W decay angles.

For the $161~\GeV$ W-pair data, only the total W-pair cross section is
used~\cite{l3-111}, while at higher centre-of-mass energies also
distributions in phase-space angles are analysed.  For charged
leptons, the sign of their electric charge determines whether they are
fermions or antifermions.  For hadronic jets, the flavour and charge
of the original quark is not measured.  Thus, a two-fold ambiguity
arises in the decay angles of hadronically decaying W bosons.

\begin{figure}[p]
\begin{center}
\epsfig{file=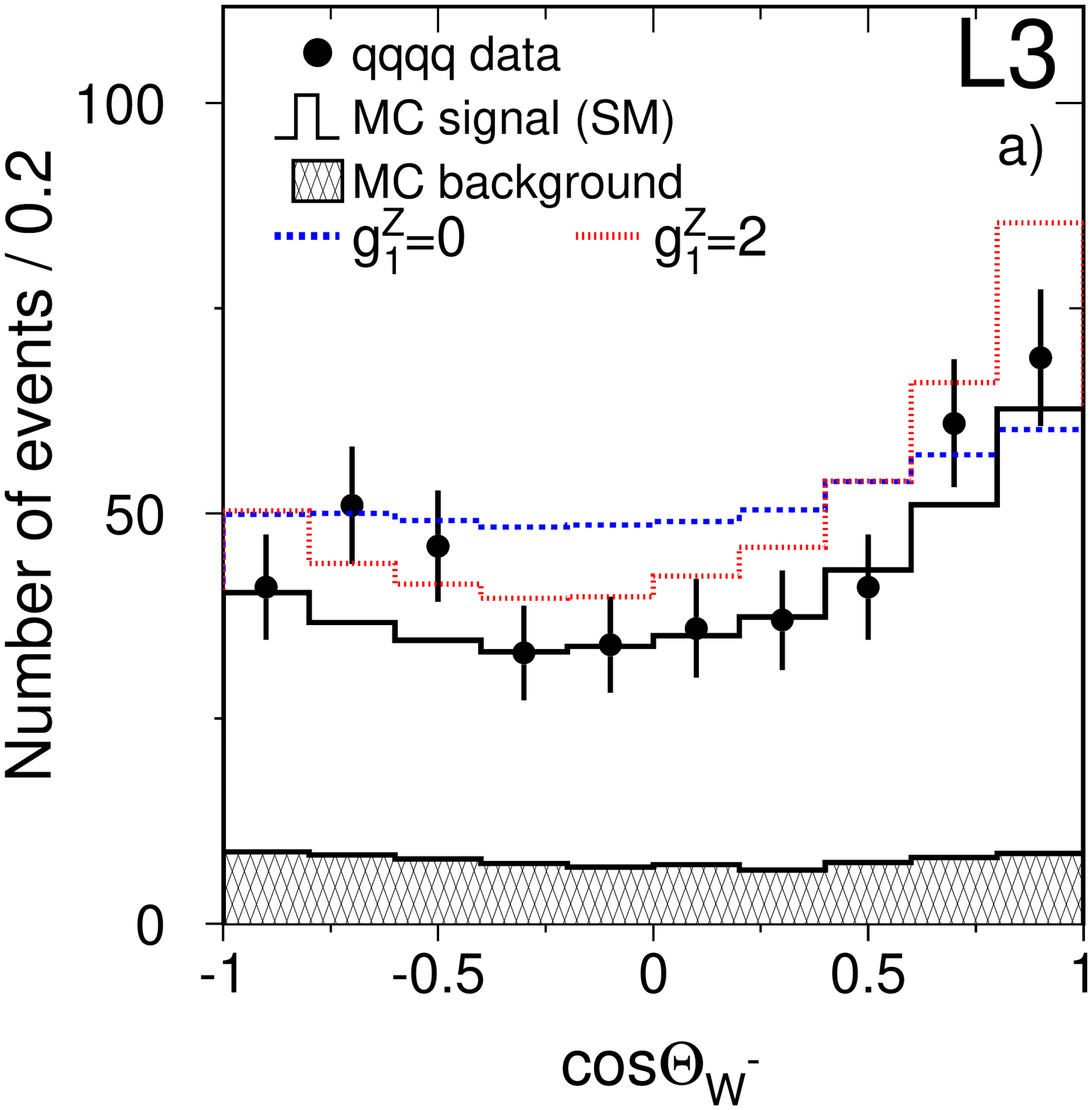,width=0.49\linewidth}\hfill
\epsfig{file=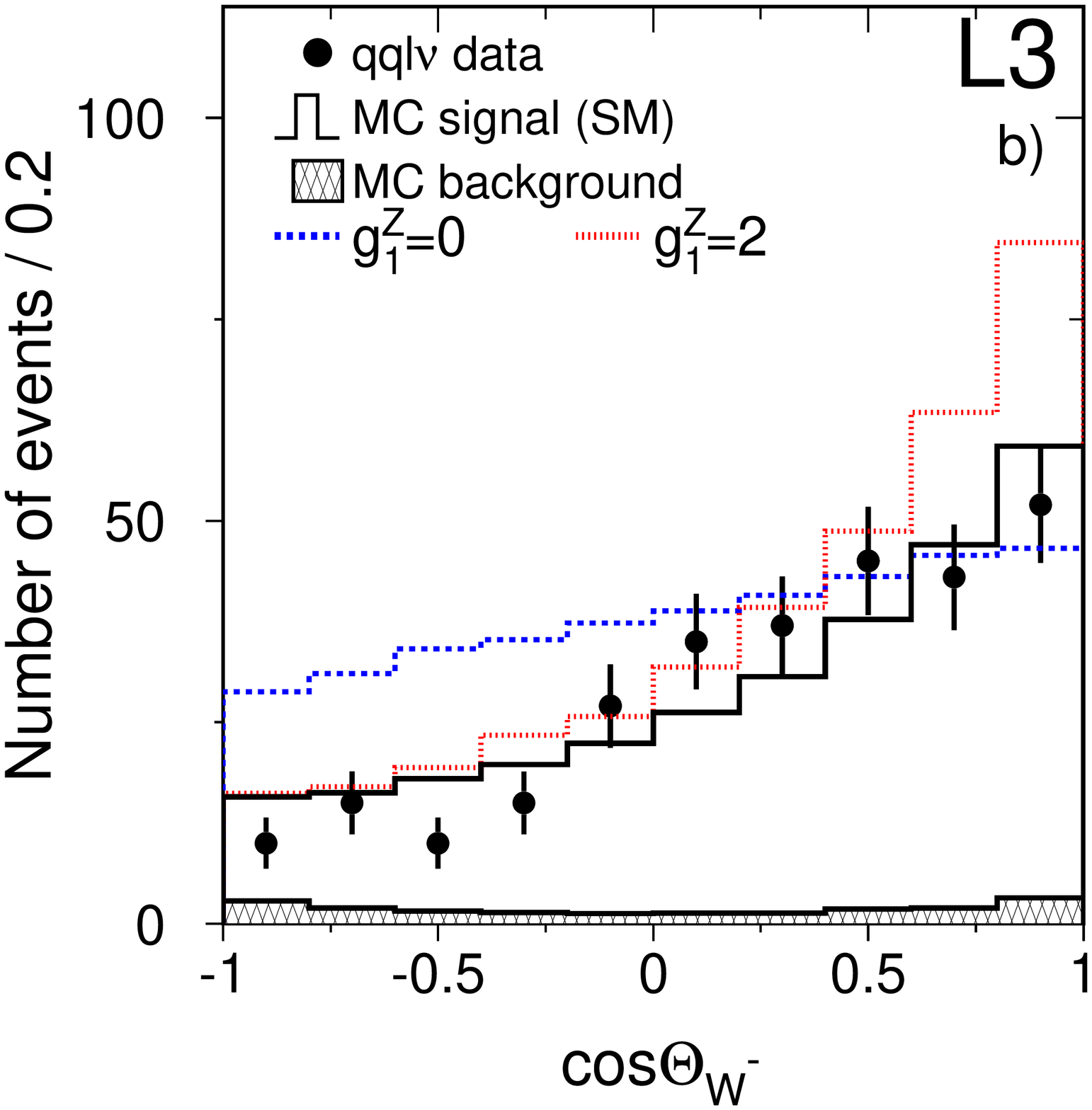,width=0.49\linewidth}\\
\hfill
\epsfig{file=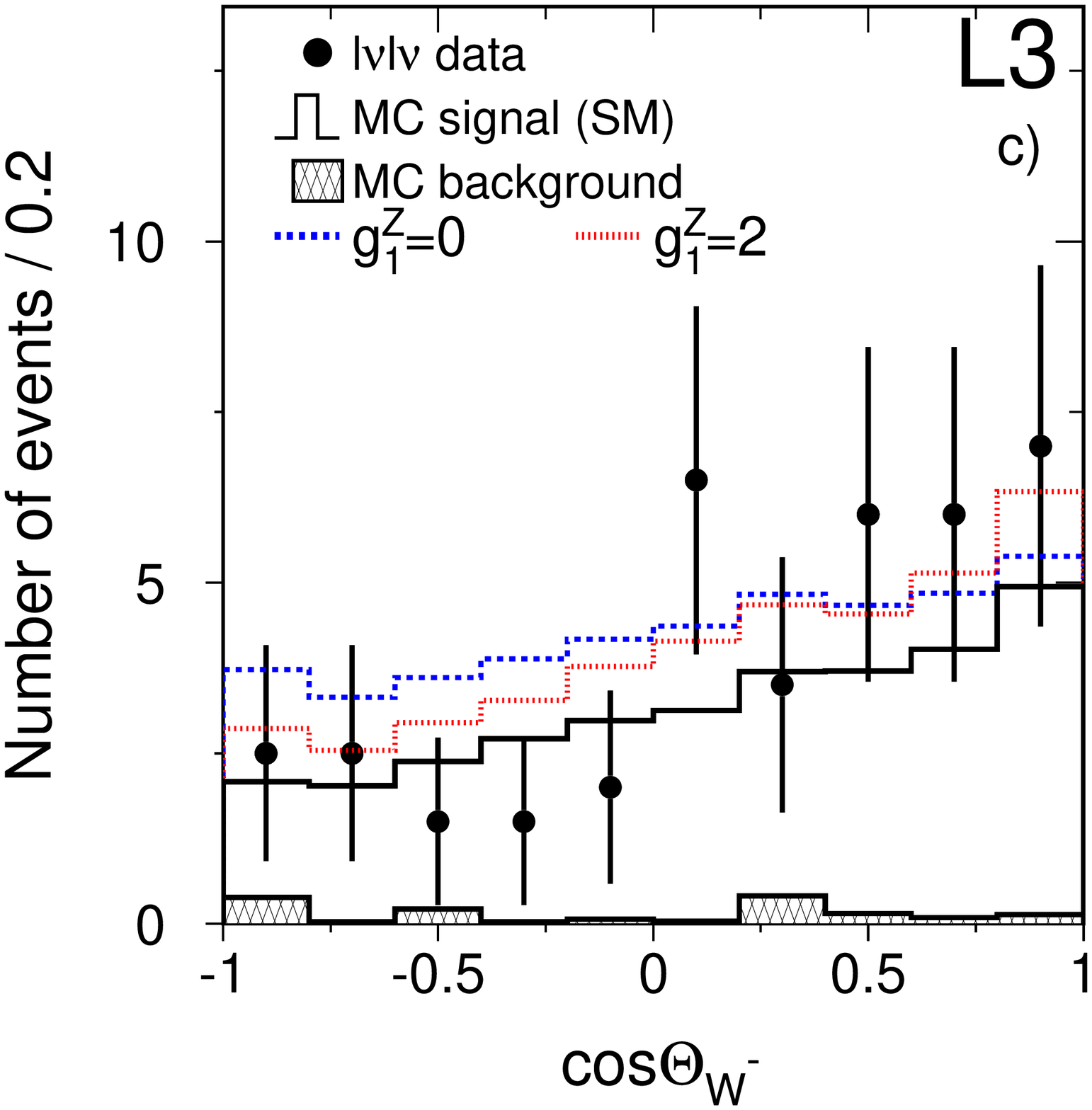,width=0.49\linewidth} 
\caption[]{Distributions of the reconstructed polar scattering angle,
  $\cos\Theta_{\PW}$, of the W$^-$ boson in W-pair events for a)
  $\QQQQ$, b) $\QQLN$, c) $\LNLN$ events. The data collected at
  $\sqrt{s}=183~\GeV$ are shown, together with the expectations for
  the SM ($\giZ=1$), and for anomalous TGCs ($\giZ=0$ or $2$).  For
  $\LNLN$ events, both solutions enter with a weight of 0.5. }
\label{fig:wwctw}
\end{center}
\end{figure}

In $\QQQQ$ events the charges of the two pairs of jets are evaluated,
based on a jet-charge technique~\cite{l3-130}, to assign positive and
negative charge to the reconstructed W bosons.  For events with
correctly paired jets, the W charge assignment is found to be correct
in 69\% of all selected $\QQQQ$ events at $\sqrt{s} = 183~\GeV$.  The
distribution of the W$^-$ polar scattering angle, $\Theta_{\PW}$,
shown in Figure~\ref{fig:wwctw}a, and the total cross section are used
for the determination of TGCs.

\begin{figure}[p]
\begin{center}
\epsfig{file=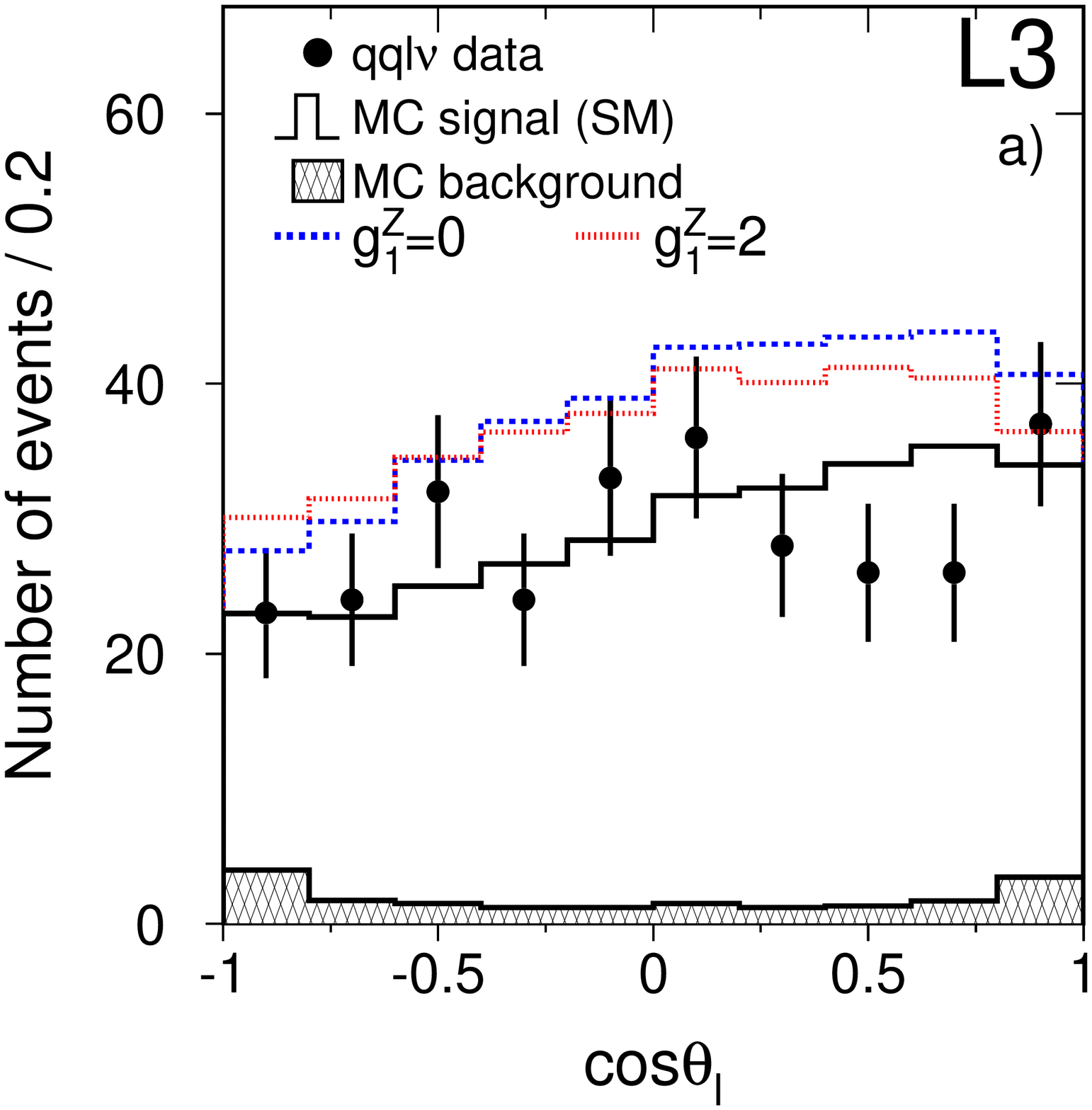,width=0.6\linewidth}
\vskip -1cm
\epsfig{file=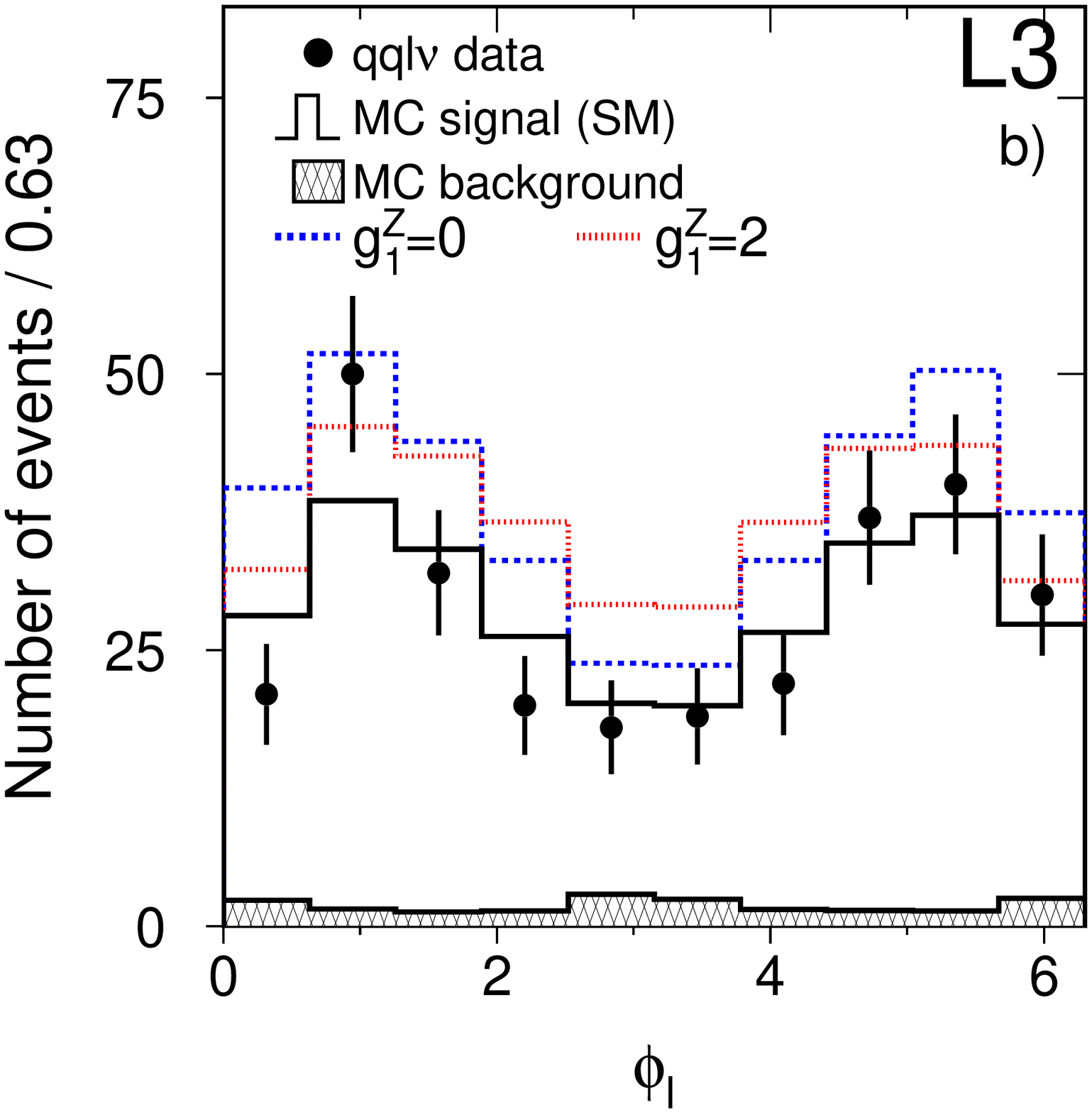,width=0.6\linewidth}
\vskip -0.5cm
\caption[]{Distributions of the reconstructed W decay angles in
  $\QQLN$ W-pair events, a) $\cos\theta_\ell$ and b) $\phi_\ell$.  The
  data collected at $\sqrt{s}=183~\GeV$ are shown, together with the
  expectations for the SM ($\giZ=1$), and for anomalous TGCs,
  ($\giZ=0$ or $2$).  The $\phi_\ell$ distribution for W$^-$ decays is
  shifted by $\pi$ in order to have the same $\phi_\ell$ distribution
  for W$^-$ and W$^+$ decays.}
\label{fig:wwstar}
\end{center}
\end{figure}

In $\QQLN$ events the W charge assignment is given by the charge of
the lepton.  The total cross section and the threefold differential
distribution in the W$^-$ polar scattering angle, $\Theta_{\PW}$, and
the two decay angles of the leptonically decaying W boson,
$\theta_\ell$ and $\phi_\ell$, are used for the determination of TGCs.
The corresponding three one-dimensional projections are shown in
Figures~\ref{fig:wwctw}b and~\ref{fig:wwstar}.

In $\LNLN$ events two unmeasured neutrinos are present.  Knowing the
momentum and charge of both charged leptons, it is possible to
calculate the polar scattering angle of the W$^-$ boson up to a
twofold ambiguity arising from the solutions of a quadratic equation.
This requires imposing energy-momentum conservation, fixing the masses
of the two W bosons to $\MW=80.41~\GeV$~\cite{PDG98}, and neglecting
photon radiation.  Both solutions are considered, each weighted by a
factor of 0.5.  In two cases the true W polar angle is not
reconstructed: (1) if one or both of the leptons is a $\tau$, the
visible lepton energy entering the calculation is not the energy of
the produced $\tau$ lepton; (2) if the two solutions are complex,
which occurs in 23\% of the selected $\LNLN$ events, their imaginary
parts are dropped.  However, the distributions still show sensitivity
to TGCs.  The distribution of the polar scattering angle,
$\Theta_{\PW}$, shown in Figure~\ref{fig:wwctw}c, and the total cross
section are used for the determination of TGCs.

\subsection{Single-W Events}
\label{sec:1W}

If in the process $\EE\rightarrow\PW\EN$ the final state electron is
scattered at low polar angles, it escapes detection along the beam
pipe.  Only the decay products of a single W boson are observed.

Leptonic single-W events, where the W boson decays into a
lepton-neutrino pair, are selected by requiring a single charged
lepton, either electron, muon or $\tau$ jet, without any other
activity in the detector~\cite{l3-112+150}.  Only the total cross
section is used in the determination of TGCs.

Hadronic single-W events, where the W boson decays into a quark pair,
are selected with a neural network approach~\cite{l3-112+150}.
After a preselection, several kinematic variables are fed into a
neural network trained to separate the hadronic single-W events from
the dominating background of $\QQLN$ W-pair events.  The total cross
section and the distribution of the neural-network output variable are
used in the determination of TGCs.

\begin{figure}[p]
\begin{center}
\epsfig{file=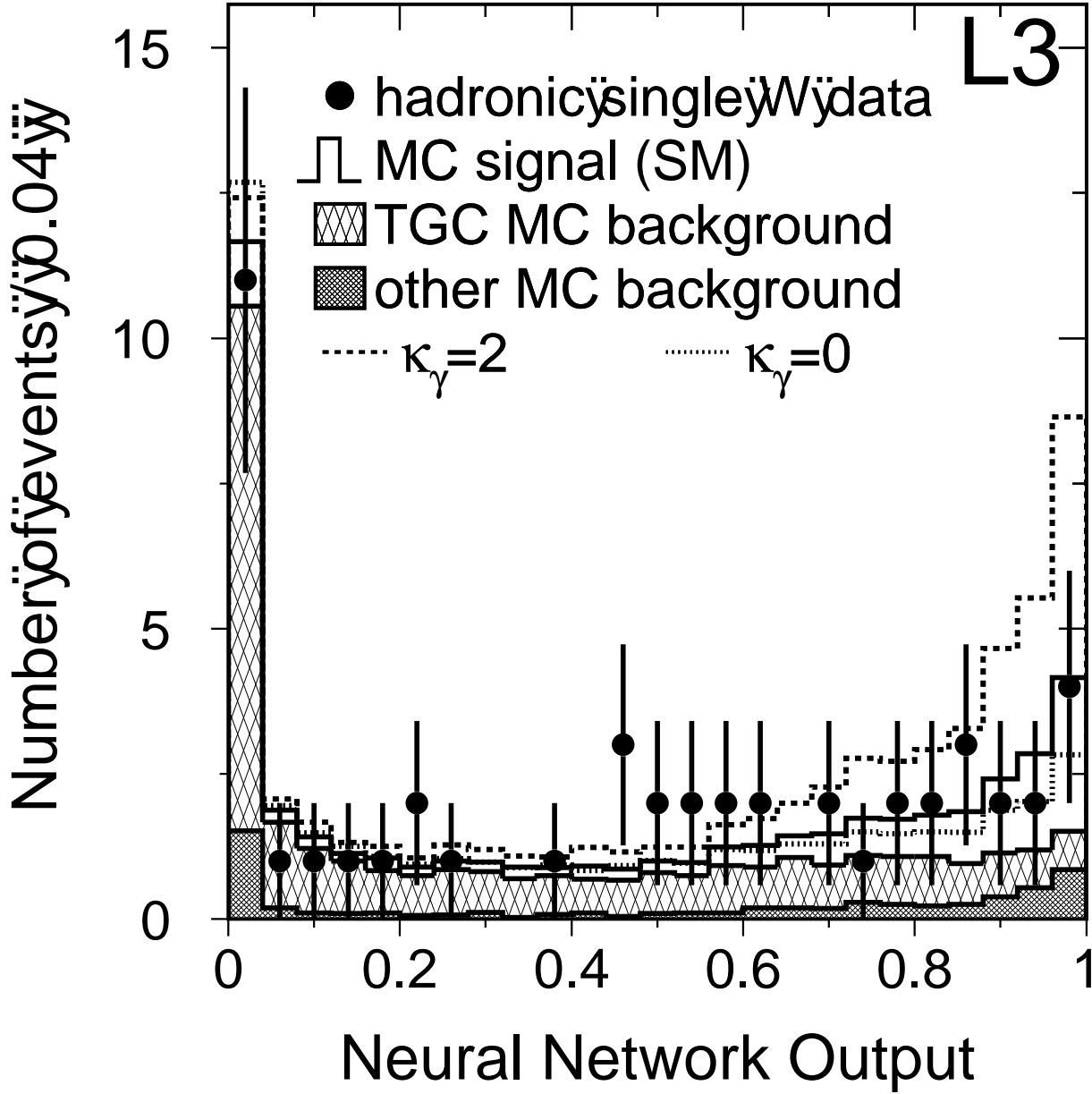,width=\linewidth} 
\caption[]{Distribution of the output of the neural network used in
  the selection of hadronic single-W events.  The data collected at
  $\sqrt{s}=183~\GeV$ are shown, together with the expectations for
  the SM ($\kg=1$), and for anomalous TGCs ($\kg=0$ or $2$).  The
  background expectation is separated into TGC-dependent W-pair
  background and other background independent of TGCs. }
\label{fig:1w}
\end{center}
\end{figure}

About a third of the hadronic single-W events are also selected by the
$\QQLN$ W-pair selections, see Table~\ref{tab:events-sub}, mainly
$\QQTN$ events.  In order to avoid double counting, such events are
considered in the W-pair sample only.  This reduces the sensitivity of
the single-W sample to TGCs as compared to the sensitivity obtained in
our published single-W TGC analysis~\cite{l3-112+150}.  However, in
combination with the W-pair signal the overall sensitivity is expected
to be better than that achieved when removing the duplicate events
from the $\QQLN$ W-pair samples.  The resulting distribution of the
output of the neural network is shown in Figure~\ref{fig:1w}.

\subsection{Single-Photon Events}
\label{sec:1g}

Single-photon events are selected by requiring one energy deposition
above $5~\GeV$ inside a polar angular range from 14$^\circ$ to
166$^\circ$ in the electromagnetic calorimeter with an electromagnetic
shower shape and without any other activity in the
detector~\cite{l3-124+160}.  Because of azimuthal symmetry, there
are two relevant observables for single-photon events, the energy and
the polar angle of the single photon.  

\begin{figure}[p]
\begin{center}
\epsfig{file=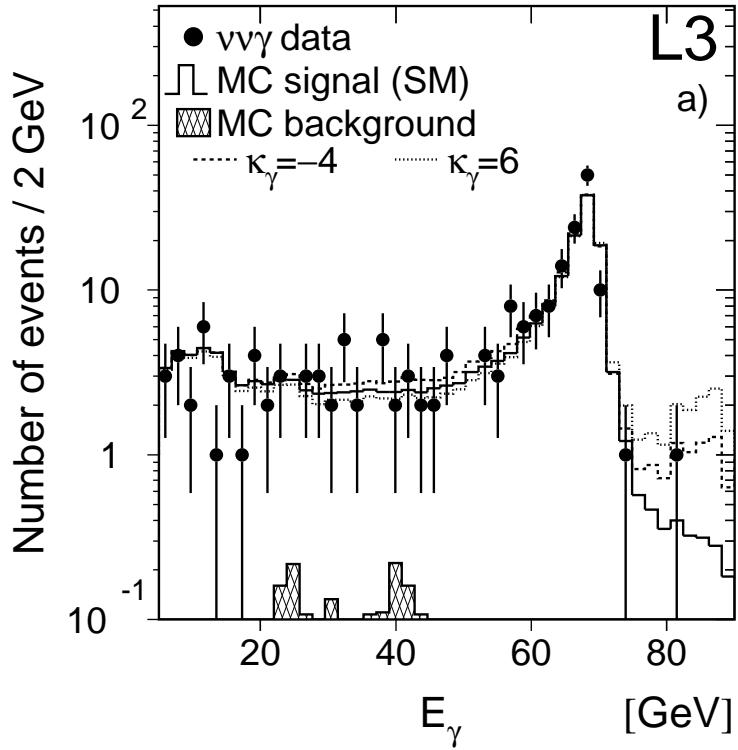,width=0.6\linewidth}
\vskip -1cm
\epsfig{file=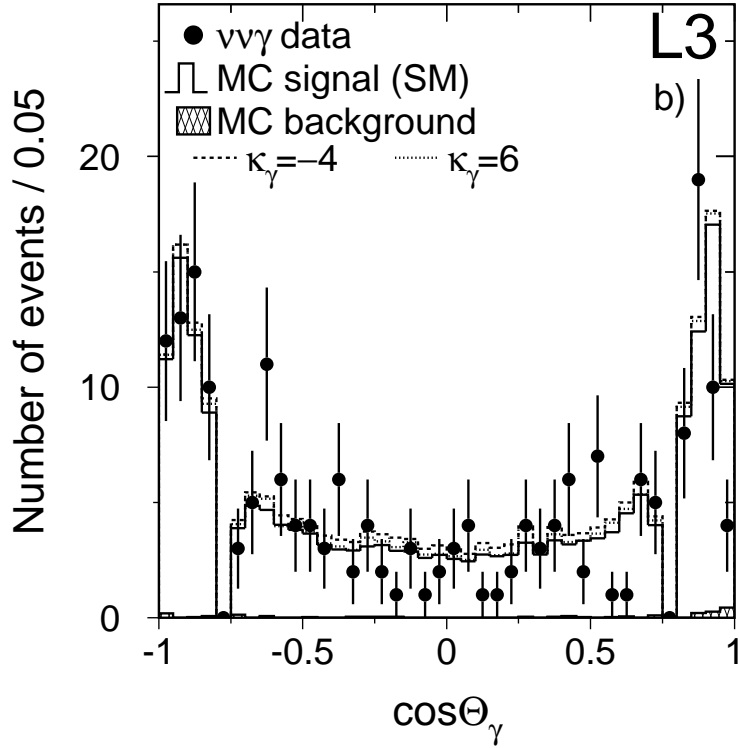,width=0.6\linewidth}
\vskip -0.5cm
\caption[]{Distributions of a) the energy, $E_\gamma$,  and b) the
  polar angle, $\Theta_\gamma$, of the photon in single-photon events.
  The data collected at $\sqrt{s}=183~\GeV$ are shown, together with
  the expectations for the SM ($\kg=1$), and for anomalous TGCs
  ($\kg=-4$ or $6$).  The main sensitivity of single-photon events to
  TGCs occurs at large photon energies. }
\label{fig:1g}
\end{center}
\end{figure}

The total cross section and the shape of the twofold differential
distribution in these two observables are used in the determination of
TGCs.  The corresponding two one-dimensional projections are shown in
Figure~\ref{fig:1g}.  The main sensitivity of single-photon events to
TGCs occurs at high photon energies above those corresponding to the
radiative return to the Z.

\section{Fitting Method}
\label{sec:fit}

The fitting procedure uses the maximum likelihood method to extract
values and errors for one or more of the TGCs denoted as $\Psi$ for
short in the following.  It is similar to the fitting procedure used
in our analysis for the mass and width of the W boson~\cite{l3-171}.

For each data event, the fit considers the set of values of the
reconstructed observables, $\Omega$, as discussed in
Sections~\ref{sec:WW}, \ref{sec:1W} and~\ref{sec:1g}.  The likelihood
is the product of the normalised differential cross section,
$L(\Omega,\Psi)$, for all data events, calculated as a function of the
TGCs $\Psi$ to be fitted.  For a given final state $i$, one has:
\begin{eqnarray}
L_{i}(\Omega_{i},\Psi) & = & 
\frac{1}{\sigma_{i}^{\SIG}(\Psi)+\sigma_{i}^{\BG}(\Psi)}
\left[
      \frac{\d\sigma_{i}^{\SIG}(\Omega_{i},\Psi)}{\d\Omega_{i}} +
      \frac{\d\sigma_{i}^{\BG} (\Omega_{i},\Psi)}{\d\Omega_{i}}
\right]
\,,
\end{eqnarray}
where $\sigma^{\SIG}_{i}$ and $\sigma_{i}^{\BG}$ are the accepted
signal and background cross sections.  For the background which is
independent of the TGCs $\Psi$, the total and differential cross
sections are taken from Monte Carlo simulations.

For values $\Psi_{\fit}$ varied during the fitting procedure, the
$\Psi$-dependent total and differential signal and background cross
sections are determined by a reweighting procedure applied to Monte
Carlo events originally generated with TGC values $\Psi_{\gen}$.  The
event weights $R_{i}$ are calculated as the ratio:
\begin{eqnarray}
R_{i}(p_n,\Psi_{\fit},\Psi_{\gen}) & = & 
\frac
{\left|{\cal M}_{i}(p_n,\Psi_{\fit})\right|^2}
{\left|{\cal M}_{i}(p_n,\Psi_{\gen})\right|^2}\,,
\end{eqnarray}
where ${\cal M}_{i}$ is the matrix element of the considered final
state $i$ evaluated for the generated four-momenta $p_n$ including
radiated photons.  For W-pair and single-W events the matrix elements
as implemented in the EXCALIBUR~\cite{EXCALIBUR} event generator are
used, which include all relevant tree-level Feynman diagrams
contributing to a given four-fermion final state. The reweighting
procedure is checked by comparisons with GENTLE~\cite{gentle} and
GRC4F~\cite{grace} cross section predictions, especially important in
the case of single-W production as GRC4F includes the effects of
fermion masses.  In the case of single-photon production the matrix
element as implemented in KORALZ~\cite{KORALZ} is used.  The
reweighting procedure is tested by comparing reweighted distributions
to distributions generated with NNGPV~\cite{nngpv} at various values
of TGCs. In all cases good agreement is observed.

The total accepted cross section for a given set of parameters
$\Psi_{\fit}$ is then:
\begin{eqnarray}
  \sigma_{i}(\Psi_{\fit}) & = & 
  \frac{\sigma^{\gen}_{i}}{N^{\gen}_{i}} \cdot
  \sum_{j}R_{i}(j,\Psi_{\fit},\Psi_{\gen})\,,
\end{eqnarray}
where $\sigma^{\gen}_{i}$ denotes the cross section corresponding to
the total Monte Carlo sample containing $N^{\gen}_{i}$ events. The sum
extends over all accepted Monte Carlo events $j$.  The accepted
differential cross section in reconstructed quantities $\Omega_{i}$ is
determined by averaging Monte Carlo events inside a box in
$\Omega_{i}$ around each data event\cite{BOXMETHOD}:
\begin{eqnarray}
\frac{\d\sigma_{i}(\Omega_{i},\Psi_{\fit})}{\d\Omega_{i}}
& = & \frac{\sigma^{\gen}_{i}}{N^{\gen}_{i}} \cdot
      \frac{1}{\Delta^{\Omega}_{i}}
      \sum_{j \epsilon 
               \Delta^{\Omega}_{i}}R_{i}(j,\Psi_{\fit},\Psi_{\gen})
\,,
\end{eqnarray}
where $\Delta^\Omega_{i}$ is the volume of the box and the sum extends
over all accepted Monte Carlo events $j$ inside the box.  This takes
$\Omega_{i}$-dependent detector effects and $\Psi$-dependent
efficiencies and purities properly into account.  The boxes are
constructed in such a way that the Monte Carlo events in the box are
the 200 events closest to the data point.  The box size in each of the
observables is proportional to the experimental resolution in those
variables.

Extended maximum likelihood fits are performed, including the overall
normalisations according to the measured total cross sections.  The
likelihood is multiplied by the Poissonian probability to obtain the
numbers of events observed in the data, given the luminosity and the
expectations for the accepted signal and background cross sections,
$\sigma_{i}^{\SIG}(\Psi_{\fit})$ and $\sigma_{i}^{\BG}(\Psi_{\fit})$.

The fitting method described above determines the TGCs without any
bias as long as the Monte Carlo describes photon radiation and
detector effects such as resolution and acceptance functions
correctly.  By fitting large Monte Carlo samples, typically a hundred
times the data, the fitting procedure is tested to high accuracy.  The
fits reproduce well the values of the TGCs of the large Monte Carlo
samples being fitted, varied in a range corresponding to three times
the error expected for the size of the data samples analysed.  Also,
the fit results do not depend on the values of the TGCs $\Psi_{\gen}$
of the Monte Carlo sample subjected to the reweighting procedure.

The reliability of the statistical errors as given by the fit is
tested by fitting for each final state several hundred small Monte
Carlo samples, each the size of the data samples.  The width of the
distribution of the fitted central values agrees well with the mean of
the distribution of the fitted errors.

\section{Triple-Gauge-Boson Couplings of the W Boson}
\label{sec:tgcs}

\begin{figure}[p]
\begin{center}
\epsfig{file=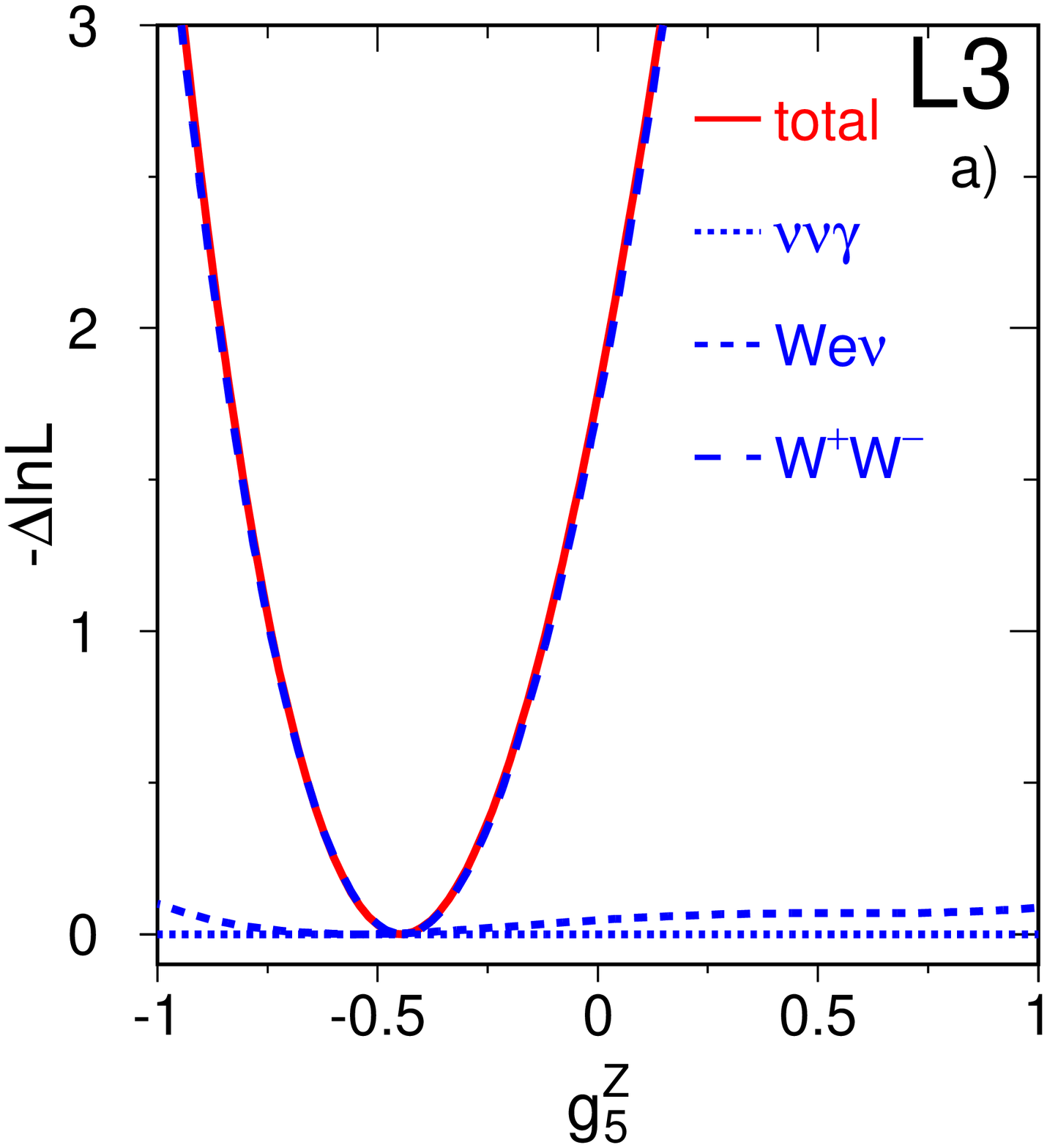,width=0.49\linewidth}\hfill
\epsfig{file=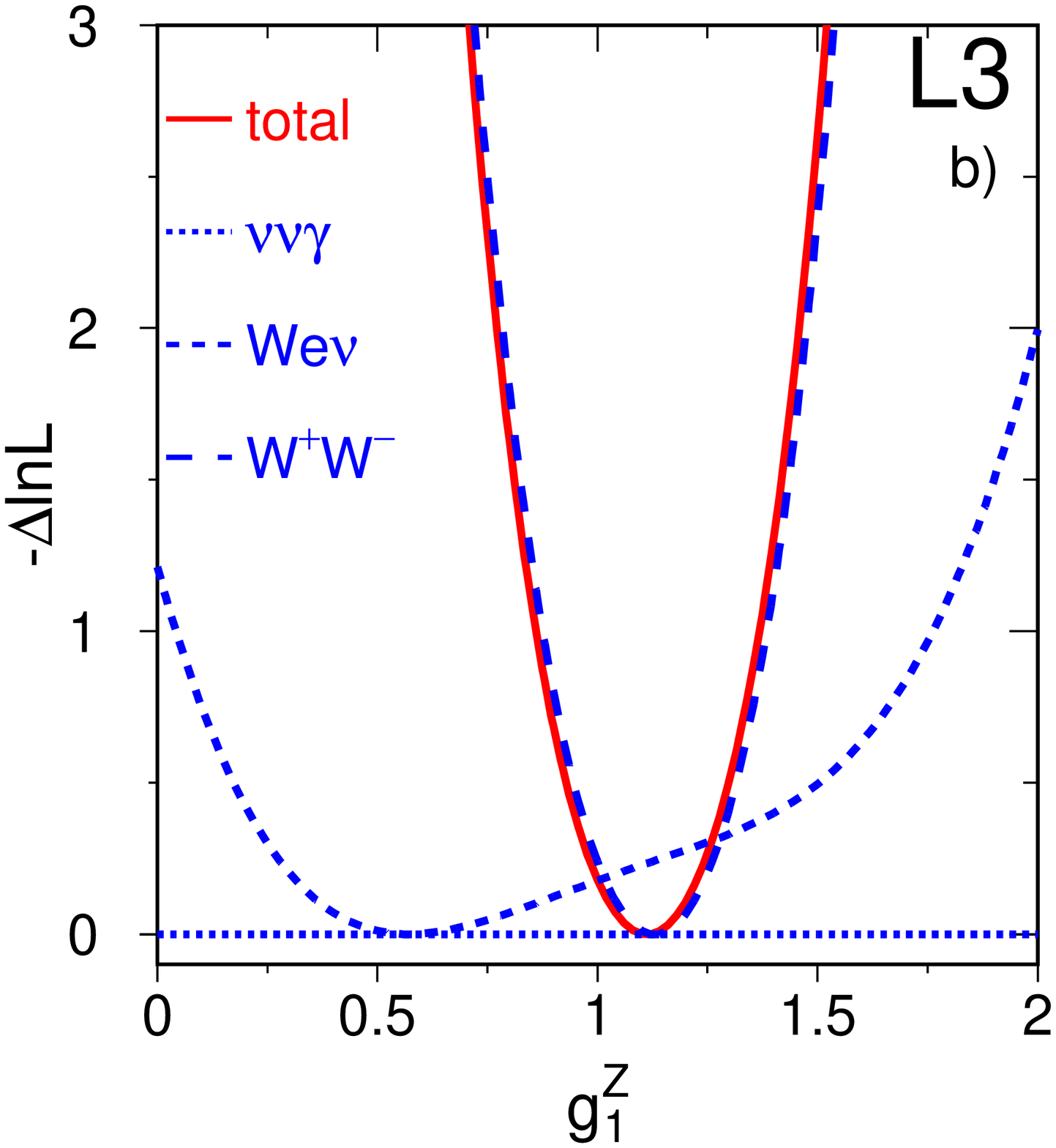,width=0.49\linewidth}\\
\epsfig{file=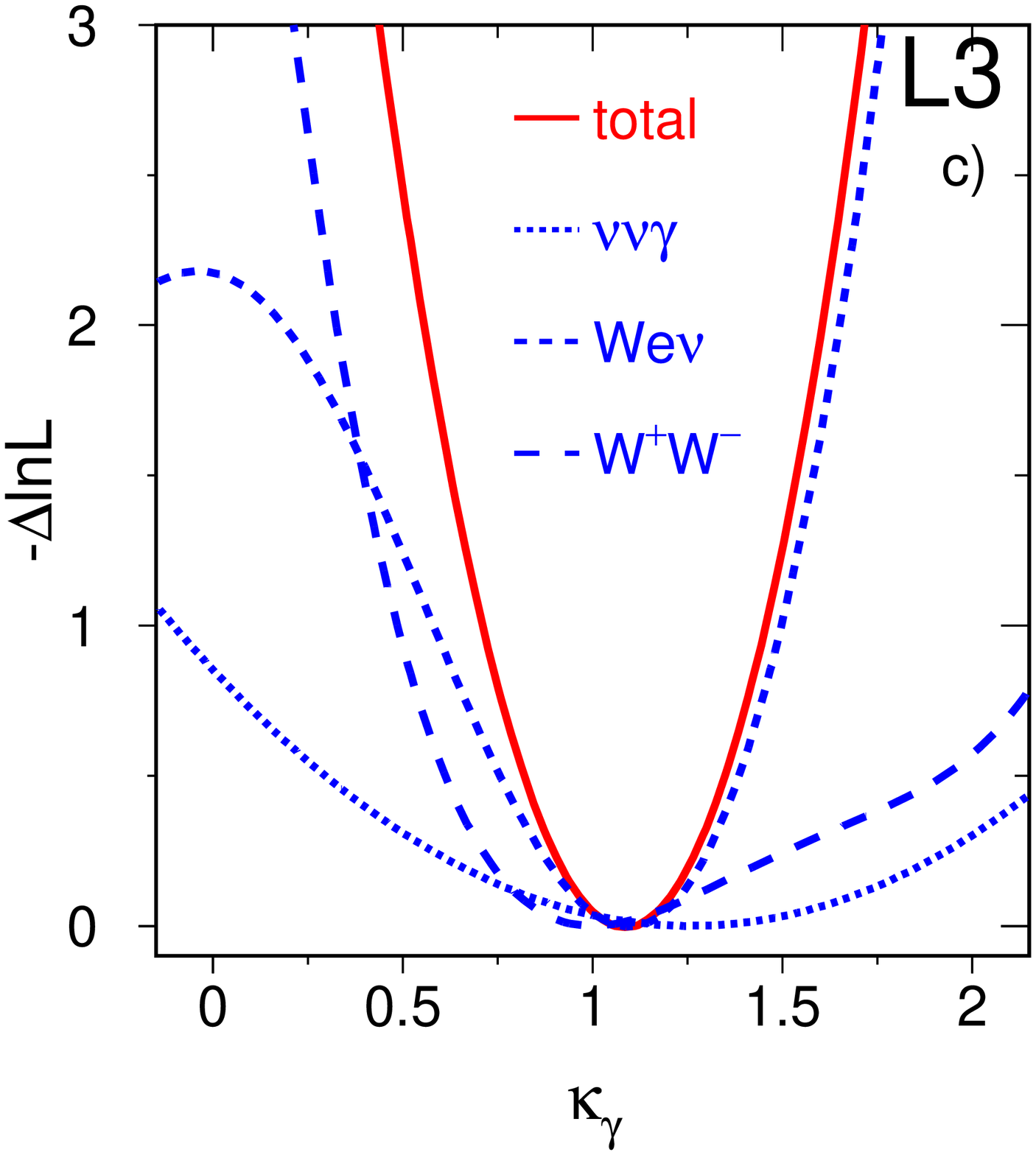,width=0.49\linewidth}\hfill
\epsfig{file=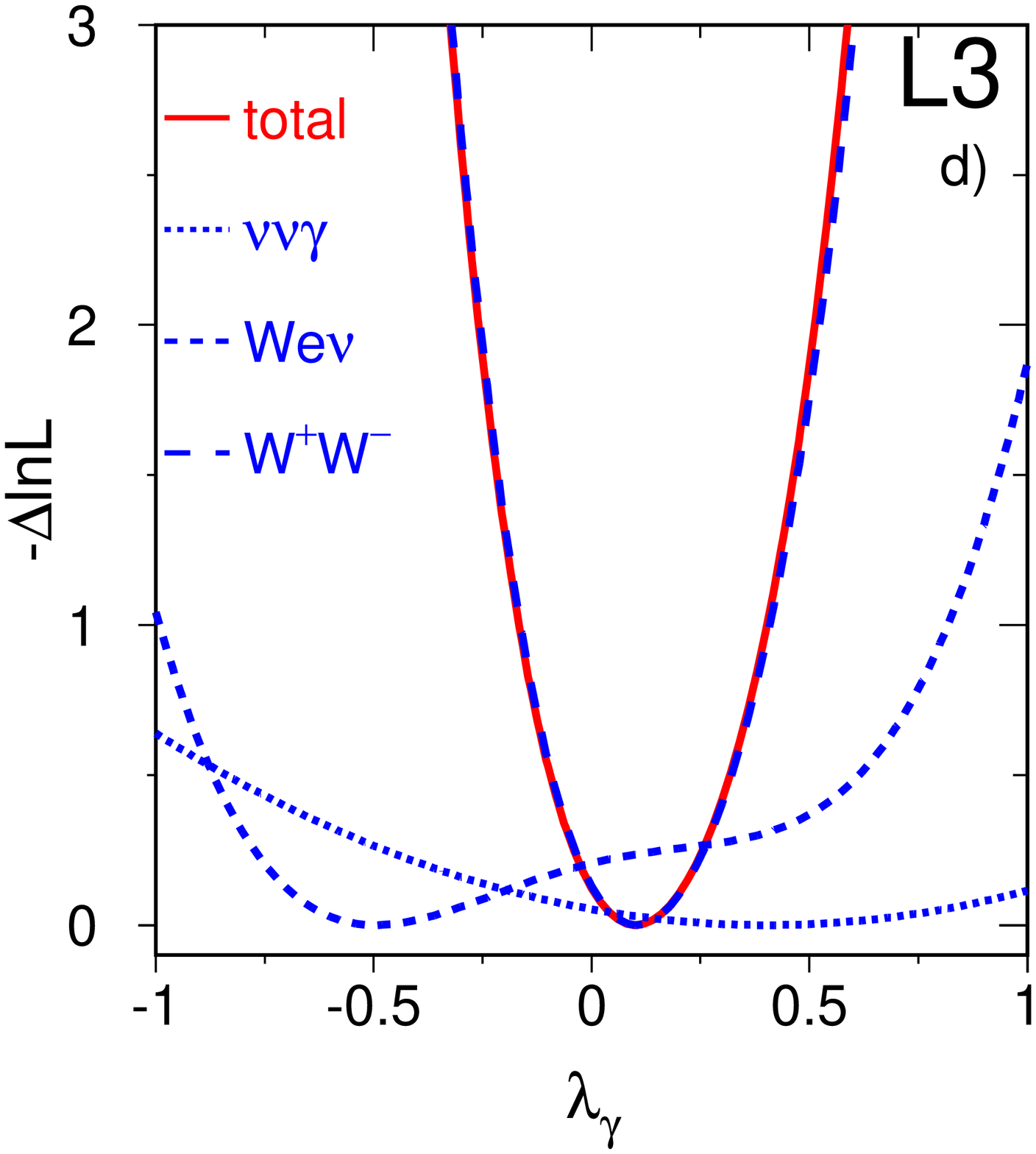,width=0.49\linewidth}
\caption[]{
  Negative log-likelihood functions (statistical errors only) for
  one-parameter fits to the TGCs a) $\gvZ$, b) $\giZ$, c) $\kg$ and d)
  $\Lg$.  The constraints $\DkZ = \Delta\giZ - \Dkg \tan^2 \theta_w$
  and $\LZ = \Lg$ are imposed. }
\label{fig:ac-1dlnl}
\end{center}
\end{figure}

In W-pair, single-W and single-photon production, the
triple-gauge-boson vertices are tested at different momentum-transfer
scales $Q^2$.  For each TGC investigated, the results derived at
$Q^2=s$ from W-pair production, at $Q^2=\MW^2$ from single-W
production, and at $Q^2=0$ from single-photon production are in good
agreement with each other, as shown in Table~\ref{tab:ac-res-1} and
Figure~\ref{fig:ac-1dlnl}.  No $Q^2$ dependence is observed.  Combined
results, obtained by adding the individual log-likelihood functions as
shown in Figure~\ref{fig:ac-1dlnl}, are reported in
Tables~\ref{tab:ac-res-1} and~\ref{tab:ac-res-2}.  Compared to the
current level of statistical accuracy, the $Q^2$ dependence of the
TGCs expected in the SM is negligible~\cite{Hagiwara87} and is thus
ignored in the combination.  As a cross check, the method of optimal
observables~\cite{optiobs1,optiobs2} is used in the case of $\QQLN$
W-pair events at $\sqrt{s}=183~\GeV$.  Compatible results are
obtained.

The statistical errors on TGCs observed in W-pair production are
larger than the expected statistical errors which are also reported in
Table~\ref{tab:ac-res-1}.  The large total cross section measured in
the $\QQQQ$ W-pair process at $\sqrt{s}=183~\GeV$~\cite{l3-155} and
the quadratic dependence of the theoretical W-pair cross section on
the TGCs cause the negative log-likelihood functions for this final
state to exhibit a two-minima structure.  Thus the sum of all
log-likelihood functions has a smaller curvature than expected with SM
cross sections.  Expected errors calculated based on the observed
cross sections agree well with the observed errors.

Multi-parameter fits of TGCs are also performed, which allow for a
more model-independent general interpretation of the data.  Fits to
two of the three C- and P-conserving TGCs $\giZ$, $\kg$, and $\Lg$,
keeping the third fixed at its SM value, as well as a fit to all three
of these TGCs are performed.  In each case the constraints $\gvZ=0$,
$\DkZ = \Delta g_1^{\mathrm{Z}} - \Dkg \tan^2 \theta_w$ and $\LZ =
\Lg$ are imposed.  A three-parameter fit to the TGCs $\giZ$, $\kg$ and
$\kZ$ is also performed, now imposing the constraints $\gvZ=\Lg=\LZ=0$
instead.  For the four-parameter fit to $(\kg,\Lg,\kZ,\LZ)$, the
constraints between the $\gamma$WW and the ZWW couplings are removed
while the constraints $\giZ=1$ and $\gvZ=0$ are imposed.  The
numerical results of all multi-parameter fits to TGCs including the
correlation matrices are reported in Table~\ref{tab:ac-res-2}.

\begin{figure}[p]
\begin{center}
\epsfig{file=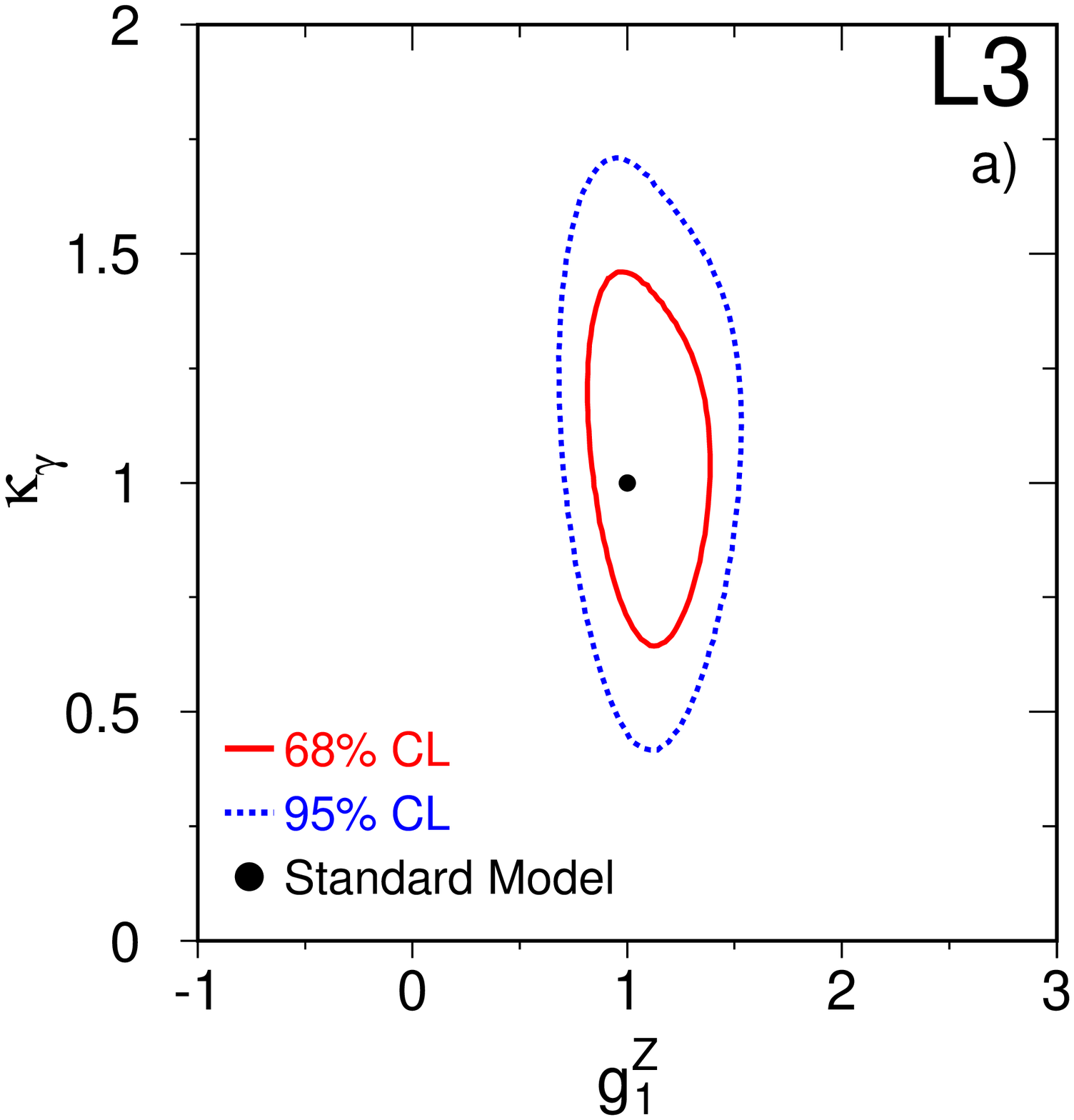,width=0.49\linewidth}\hfill
\epsfig{file=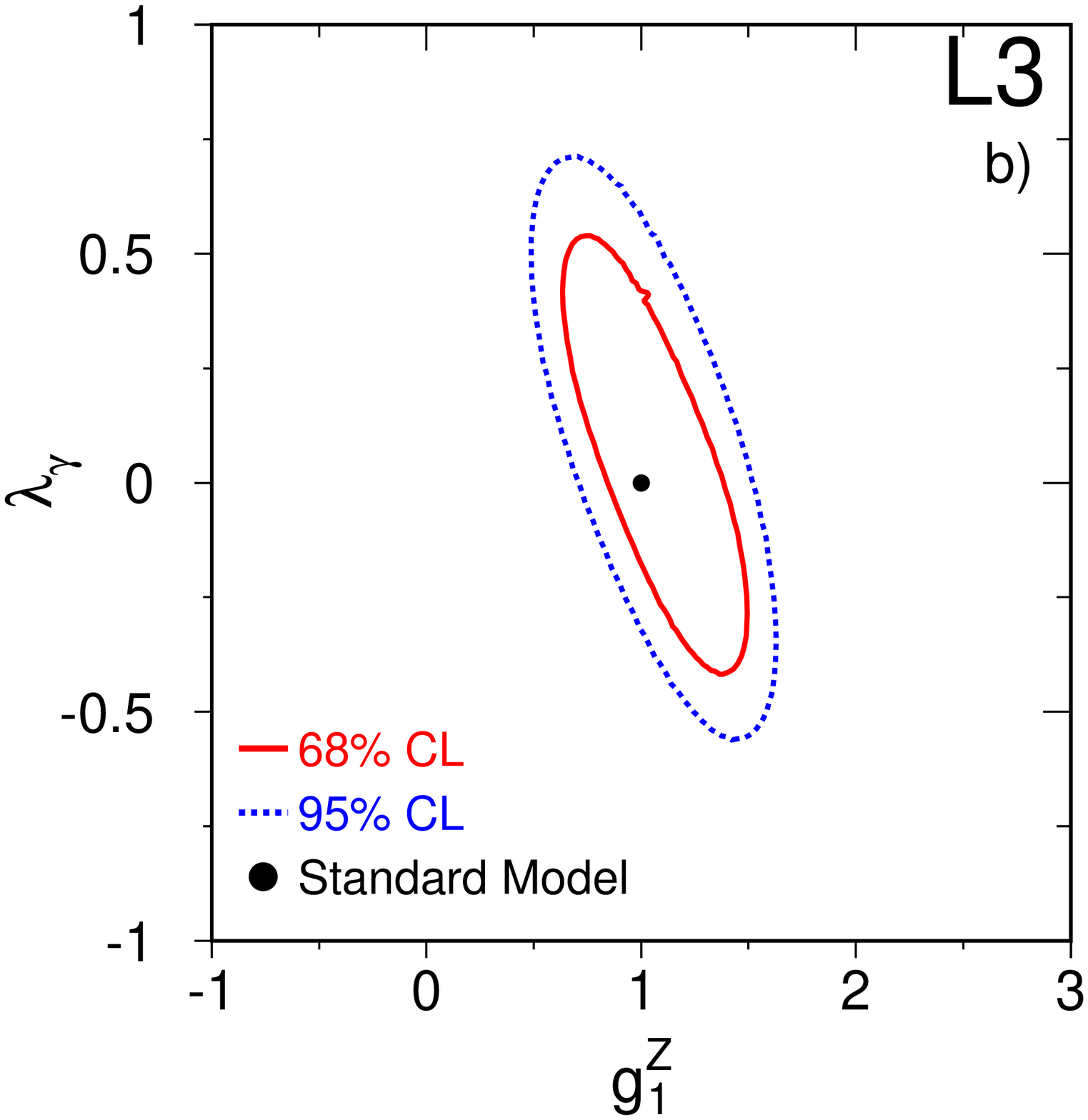,width=0.49\linewidth}\\
\epsfig{file=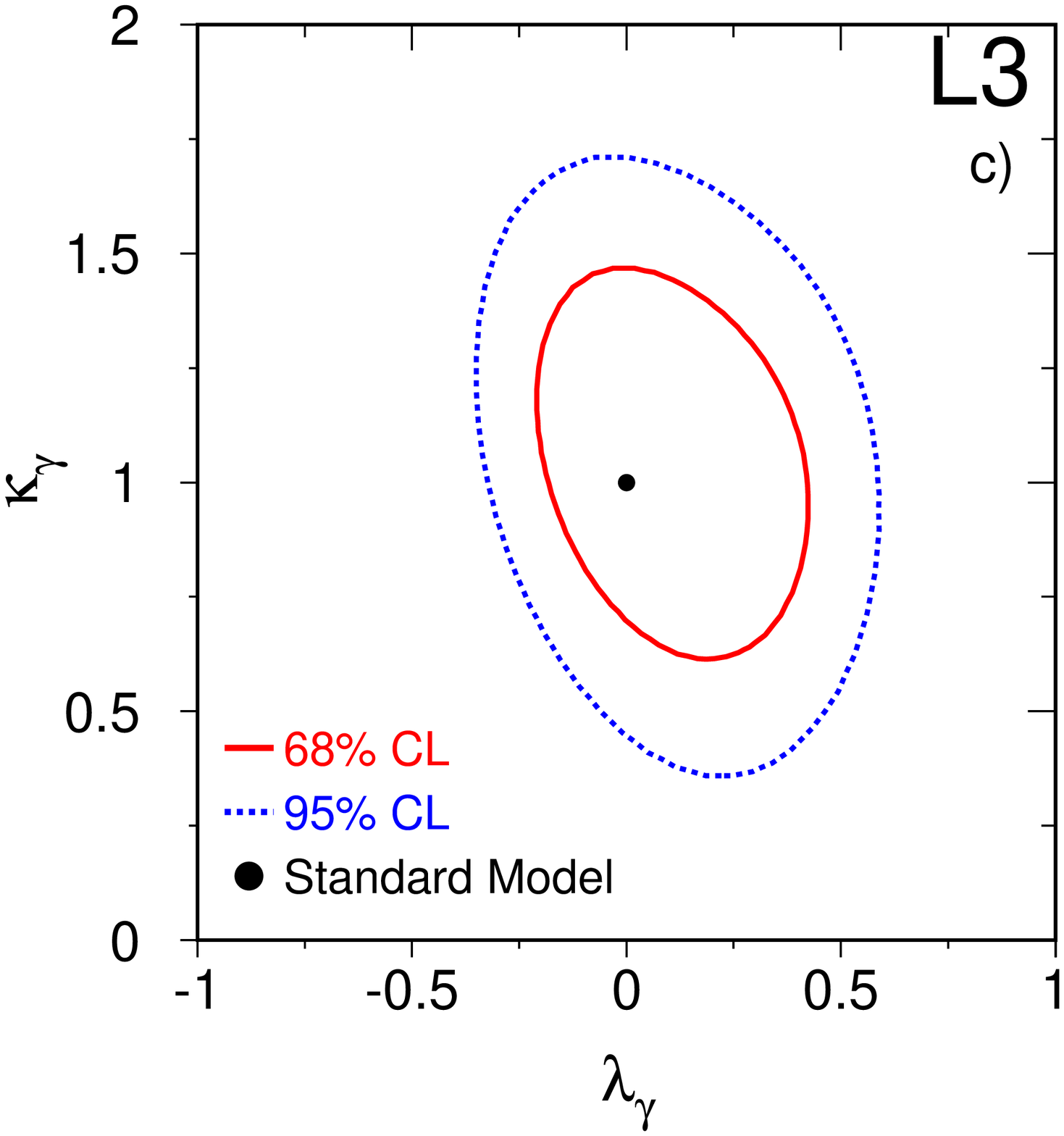,width=0.49\linewidth}
\caption[]{
  Contour curves of 68\% and 95\% probability for the two-parameter
  fits to the TGCs a) $\giZ$ and $\kg$ with $\Lg=0$, b) $\giZ$ and
  $\Lg$ with $\kg=1$, c) $\kg$ and $\Lg$ with $\giZ=1$.  The
  constraints $\gvZ=0$, $\DkZ=\Delta\giZ-\Dkg\tan^2\theta_w$ and
  $\LZ=\Lg$ are imposed.
}
\label{fig:ac-ndlnl}
\end{center}
\end{figure}

As an example, the contour curves of 68\% and 95\% probability derived
from fits to two of the three C- and P-conserving TGCs $\giZ$, $\kg$,
and $\Lg$, keeping the third fixed at its SM value, are shown in
Figure~\ref{fig:ac-ndlnl}.  The contour curves correspond to a change
in log-likelihood with respect to its minimum of 1.15 and 3.0,
respectively.

\subsection{Systematic Errors}
\label{sec:syst}

The sources of systematic errors considered include those studied for
the W-boson mass and width analysis~\cite{l3-171}: LEP energy,
initial- and final-state radiation, jet and lepton measurement,
fragmentation and decay, background normalisation and shape, Monte
Carlo statistics, fitting method, and in the case of the $\QQQQ$ final
state, colour reconnection and Bose-Einstein effects.  The methods
used to evaluate the effects on TGCs are identical.  The changes in
both the central value and the statistical error due to systematic
effects are taken into account.

The systematic errors on TGCs for the different models and processes
are summarised in Tables~\ref{tab:ac-syst-1} and~\ref{tab:ac-syst-2}.
In most cases the total systematic error is dominated by the
uncertainty in the experimental selection efficiencies and the
theoretical error of 2\% on the total cross section predictions.
Additional systematic effects arise due to uncertainties in the
description of the charge confusion affecting the W charge assignment.
Systematic errors due to uncertainties in the mass and total width of
the W boson are small.

\subsection{Results and Discussion}
\label{sec:results}

For the CP conserving but C- and P-violating coupling $\gvZ$, the
following result is obtained when all other TGCs are fixed to their SM
values:
\begin{eqnarray}
\gvZ & = & -0.44\amp{0.22}{0.23}\pm0.12 \,.
\end{eqnarray}
The first error is statistical and the second systematic.  The result
is in agreement with the SM expectation of $\gvZ=0$.  Imposing the
constraints $\gvZ=0$, $\DkZ = \Delta g_1^{\mathrm{Z}} - \Dkg \tan^2
\theta_w$ and $\LZ = \Lg$, three C- and P-conserving TGCs remain.  The
results of fits to the individual couplings, keeping the other two at
their SM values, are:
\begin{eqnarray}
\giZ & = & +1.11\amp{0.18}{0.19}\pm0.10 \\
\kg  & = & +1.11\amp{0.25}{0.25}\pm0.17 \\
\Lg  & = & +0.10\amp{0.20}{0.22}\pm0.10 \,.
\end{eqnarray}
The log-likelihood functions for all four one-parameter fits are shown in
Figure~\ref{fig:ac-1dlnl}.  The contribution from W-pair production
dominates in the case of $\gvZ$, $\giZ$ and $\Lg$, while the single-W
contribution is important for the constraint on $\kg$.  The hadronic
single-W samples also contribute to the constraints on the ZWW
couplings through the remaining couplings-dependent $\QQLN$ W-pair
background.  The single-photon samples constrain the $\gamma$WW
couplings only.

All single- and multi-parameter TGC results show good agreement with
the SM expectation and imply the existence of the self coupling among
the electroweak gauge bosons.  The resulting constraints on non-SM
contributions to TGCs are significantly improved with respect to our
previous analyses~\cite{l3-111,l3-130,l3-112+150}.

The measurements exclude a theory by Klein~\cite{klein}, predicting a
value of $\kg=-2$, by more than ten standard deviations.  If the W
boson were an extended object, \eg, an ellipsoid of rotation with
longitudinal radius $a$ and transverse radius $b$, its typical size
and shape would be related to the TGCs by $R_{\PW} \equiv (a+b)/2 =
(\kg+\Lg-1)/\MW$~\cite{Brodsky:1980} and $\Delta_{\PW} \equiv
(a^2-b^2)/2 = (5/4)(\kg-\Lg-1)/\MW^2$~\cite{quad1,quad2,quad3}.  The
measurements show no evidence for the W boson to be an extended
object:
\begin{eqnarray}
     R_{\PW} & = & (0.3\pm1.0)\cdot10^{-18}~\mathrm{m}   \\
\Delta_{\PW} & = & (0.3\pm3.1)\cdot10^{-36}~\mathrm{m}^2 \,,
\end{eqnarray}
with a correlation coefficient of $-0.26$.  These results establish
the pointlike nature of the W boson down to a scale of $10^{-18}$~m.

%
%
\section{Acknowledgements}

We wish to congratulate the CERN accelerator divisions for the
successful upgrade of the LEP machine and to express our gratitude for
its good performance. We acknowledge with appreciation the effort of
the engineers, technicians and support staff who have participated in
the construction and maintenance of this experiment.

\clearpage

%
%
\bibliographystyle{l3stylem}

%
%
\newpage
\typeout{   }     
\typeout{Using author list for paper 189 ONLY ONLY ONLY!!!!}
\typeout{$Modified: Fri Sep 10 08:43:14 1999 by clare $}
\typeout{!!!!  This should only be used with document option a4p!!!!}
\typeout{   }
%
%
%
%
%
%

\newcount\tutecount  \tutecount=0
\def\tutenum#1{\global\advance\tutecount by 1 \xdef#1{\the\tutecount}}
\def\tute#1{$^{#1}$}
\tutenum\aachen            
\tutenum\nikhef            
\tutenum\mich              
\tutenum\lapp              
\tutenum\basel             
\tutenum\lsu               
\tutenum\beijing           
\tutenum\berlin            
\tutenum\bologna           
\tutenum\tata              
\tutenum\ne                
\tutenum\bucharest         
\tutenum\budapest          
\tutenum\mit               
\tutenum\debrecen          
\tutenum\florence          
\tutenum\cern              
\tutenum\wl                
\tutenum\geneva            
\tutenum\hefei             
\tutenum\seft              
\tutenum\lausanne          
\tutenum\lecce             
\tutenum\lyon              
\tutenum\madrid            
\tutenum\milan             
\tutenum\moscow            
\tutenum\naples            
\tutenum\cyprus            
\tutenum\nymegen           
\tutenum\caltech           
\tutenum\perugia           
\tutenum\cmu               
\tutenum\prince            
\tutenum\rome              
\tutenum\peters            
\tutenum\salerno           
\tutenum\ucsd              
\tutenum\santiago          
\tutenum\sofia             
\tutenum\korea             
\tutenum\alabama           
\tutenum\utrecht           
\tutenum\purdue            
\tutenum\psinst            
\tutenum\zeuthen           
\tutenum\eth               
\tutenum\hamburg           
\tutenum\taiwan            
\tutenum\tsinghua          
{
\parskip=0pt
\noindent
{\bf The L3 Collaboration:}
\ifx\selectfont\undefined
 \baselineskip=10.8pt
 \baselineskip\baselinestretch\baselineskip
 \normalbaselineskip\baselineskip
 \ixpt
\else
 \fontsize{9}{10.8pt}\selectfont
\fi
\medskip
\tolerance=10000
\hbadness=5000
\raggedright
\hsize=162truemm\hoffset=0mm
\def\r{\rlap,}
\noindent

M.Acciarri\r\tute\milan\
P.Achard\r\tute\geneva\ 
O.Adriani\r\tute{\florence}\ 
M.Aguilar-Benitez\r\tute\madrid\ 
J.Alcaraz\r\tute\madrid\ 
G.Alemanni\r\tute\lausanne\
J.Allaby\r\tute\cern\
A.Aloisio\r\tute\naples\ 
M.G.Alviggi\r\tute\naples\
G.Ambrosi\r\tute\geneva\
% ...
\newpage
\begin{list}{A}{\itemsep=0pt plus 0pt minus 0pt\parsep=0pt plus 0pt minus 0pt
                \topsep=0pt plus 0pt minus 0pt}
\item[\aachen]
 I. Physikalisches Institut, RWTH, D-52056 Aachen, FRG$^{\S}$\\
 III. Physikalisches Institut, RWTH, D-52056 Aachen, FRG$^{\S}$
\item[\nikhef] National Institute for High Energy Physics, NIKHEF, 
     and University of Amsterdam, NL-1009 DB Amsterdam, The Netherlands
\item[\mich] University of Michigan, Ann Arbor, MI 48109, USA
\item[\lapp] Laboratoire d'Annecy-le-Vieux de Physique des Particules, 
     LAPP,IN2P3-CNRS, BP 110, F-74941 Annecy-le-Vieux CEDEX, France
\item[\basel] Institute of Physics, University of Basel, CH-4056 Basel,
     Switzerland
\item[\lsu] Louisiana State University, Baton Rouge, LA 70803, USA
\item[\beijing] Institute of High Energy Physics, IHEP, 
  100039 Beijing, China$^{\triangle}$ 
\item[\berlin] Humboldt University, D-10099 Berlin, FRG$^{\S}$
\item[\bologna] University of Bologna and INFN-Sezione di Bologna, 
     I-40126 Bologna, Italy
\item[\tata] Tata Institute of Fundamental Research, Bombay 400 005, India
\item[\ne] Northeastern University, Boston, MA 02115, USA
\item[\bucharest] Institute of Atomic Physics and University of Bucharest,
     R-76900 Bucharest, Romania
\item[\budapest] Central Research Institute for Physics of the 
     Hungarian Academy of Sciences, H-1525 Budapest 114, Hungary$^{\ddag}$
\item[\mit] Massachusetts Institute of Technology, Cambridge, MA 02139, USA
\item[\debrecen] Lajos Kossuth University-ATOMKI, H-4010 Debrecen, Hungary$^\P$
\item[\florence] INFN Sezione di Firenze and University of Florence, 
     I-50125 Florence, Italy
\item[\cern] European Laboratory for Particle Physics, CERN, 
     CH-1211 Geneva 23, Switzerland
\item[\wl] World Laboratory, FBLJA  Project, CH-1211 Geneva 23, Switzerland
\item[\geneva] University of Geneva, CH-1211 Geneva 4, Switzerland
\item[\hefei] Chinese University of Science and Technology, USTC,
      Hefei, Anhui 230 029, China$^{\triangle}$
\item[\seft] SEFT, Research Institute for High Energy Physics, P.O. Box 9,
      SF-00014 Helsinki, Finland
\item[\lausanne] University of Lausanne, CH-1015 Lausanne, Switzerland
\item[\lecce] INFN-Sezione di Lecce and Universit\'a Degli Studi di Lecce,
     I-73100 Lecce, Italy
\item[\lyon] Institut de Physique Nucl\'eaire de Lyon, 
     IN2P3-CNRS,Universit\'e Claude Bernard, 
     F-69622 Villeurbanne, France
\item[\madrid] Centro de Investigaciones Energ{\'e}ticas, 
     Medioambientales y Tecnolog{\'\i}cas, CIEMAT, E-28040 Madrid,
     Spain${\flat}$ 
\item[\milan] INFN-Sezione di Milano, I-20133 Milan, Italy
\item[\moscow] Institute of Theoretical and Experimental Physics, ITEP, 
     Moscow, Russia
\item[\naples] INFN-Sezione di Napoli and University of Naples, 
     I-80125 Naples, Italy
\item[\cyprus] Department of Natural Sciences, University of Cyprus,
     Nicosia, Cyprus
\item[\nymegen] University of Nijmegen and NIKHEF, 
     NL-6525 ED Nijmegen, The Netherlands
\item[\caltech] California Institute of Technology, Pasadena, CA 91125, USA
\item[\perugia] INFN-Sezione di Perugia and Universit\'a Degli 
     Studi di Perugia, I-06100 Perugia, Italy   
\item[\cmu] Carnegie Mellon University, Pittsburgh, PA 15213, USA
\item[\prince] Princeton University, Princeton, NJ 08544, USA
\item[\rome] INFN-Sezione di Roma and University of Rome, ``La Sapienza",
     I-00185 Rome, Italy
\item[\peters] Nuclear Physics Institute, St. Petersburg, Russia
\item[\salerno] University and INFN, Salerno, I-84100 Salerno, Italy
\item[\ucsd] University of California, San Diego, CA 92093, USA
\item[\santiago] Dept. de Fisica de Particulas Elementales, Univ. de Santiago,
     E-15706 Santiago de Compostela, Spain
\item[\sofia] Bulgarian Academy of Sciences, Central Lab.~of 
     Mechatronics and Instrumentation, BU-1113 Sofia, Bulgaria
\item[\korea] Center for High Energy Physics, Adv.~Inst.~of Sciences
     and Technology, 305-701 Taejon,~Republic~of~{Korea}
\item[\alabama] University of Alabama, Tuscaloosa, AL 35486, USA
\item[\utrecht] Utrecht University and NIKHEF, NL-3584 CB Utrecht, 
     The Netherlands
\item[\purdue] Purdue University, West Lafayette, IN 47907, USA
\item[\psinst] Paul Scherrer Institut, PSI, CH-5232 Villigen, Switzerland
\item[\zeuthen] DESY, D-15738 Zeuthen, 
     FRG
\item[\eth] Eidgen\"ossische Technische Hochschule, ETH Z\"urich,
     CH-8093 Z\"urich, Switzerland
\item[\hamburg] University of Hamburg, D-22761 Hamburg, FRG
\item[\taiwan] National Central University, Chung-Li, Taiwan, China
\item[\tsinghua] Department of Physics, National Tsing Hua University,
      Taiwan, China
\item[\S]  Supported by the German Bundesministerium 
        f\"ur Bildung, Wissenschaft, Forschung und Technologie
\item[\ddag] Supported by the Hungarian OTKA fund under contract
numbers T019181, F023259 and T024011.
\item[\P] Also supported by the Hungarian OTKA fund under contract
  numbers T22238 and T026178.
\item[$\flat$] Supported also by the Comisi\'on Interministerial de Ciencia y 
        Tecnolog{\'\i}a.
\item[$\sharp$] Also supported by CONICET and Universidad Nacional de La Plata,
        CC 67, 1900 La Plata, Argentina.
\item[$\diamondsuit$] Also supported by Panjab University, Chandigarh-160014, 
        India.
\item[$\triangle$] Supported by the National Natural Science
  Foundation of China.
\item[\dag] Deceased.
\end{list}
}
\vfill






\clearpage

\begin{table}[p]
\begin{center}
\renewcommand{\arraystretch}{1.2}
\begin{tabular}{| l || c | c || c | c || c | c |}
\hline                                    
$\sqrt{s}$ & \multicolumn{2}{|c||}{$183~\GeV$} 
           & \multicolumn{2}{|c||}{$172~\GeV$}  
           & \multicolumn{2}{|c| }{$161~\GeV$}      \\
\hline
Process & $N_{\mathrm{data}}$ & $N_{\mathrm{bg}}$  
        & $N_{\mathrm{data}}$ & $N_{\mathrm{bg}}$  
        & $N_{\mathrm{data}}$ & $N_{\mathrm{bg}}$  \\
\hline                                                                        
\hline
$\WW\rightarrow\LNLN$ & $\pz54$ & $\pz9.7$ & $\pz19$ & $\pz0.6$ & $\pzz2$ & $\pz0.4$ \\
$\WW\rightarrow\QQEN$ & $  112$ & $\pz6.7$ & $\pzz9$ & $\pz0.4$ & $\pzz4$ & $\pz0.2$ \\
$\WW\rightarrow\QQMN$ & $  108$ & $\pz5.7$ & $\pz12$ & $\pz2.1$ & $\pzz4$ & $\pz0.2$ \\
$\WW\rightarrow\QQTN$ & $\pz77$ & $  10.6$ & $\pzz9$ & $\pz0.3$ & $\pzz3$ & $\pz1.6$ \\
$\WW\rightarrow\QQQQ$ & $  473$ & $  81.2$ & $\pz61$ & $  12.6$ & $\pz11$ & $\pz5.1$ \\
\hline                                                                       
$\PW\EN,~\PW\rightarrow qq$ & $\pz86$ & $  72.6$ & $\pz15$ & $  10.1$ & $\pzz7$ & $\pz5.5$ \\
$\PW\EN,~\PW\rightarrow\LN$ & $\pz10$ & $\pz3.1$ & $\pzz1$ & $\pz0.4$ & $\pzz1$ & $\pz0.4$ \\
\hline                                                                       
$\NN\gamma$ & $  198$ & $\pz2.1$ & $\pz52$ & $\pz0.3$ & $\pz59$ & $\pz0.6$ \\
\hline
\end{tabular}
\caption[]{
  Number of selected data events, $N_{\mathrm{data}}$, and expected
  background events, $N_{\mathrm{bg}}$, in W-pair, single-W and
  single-photon production. }
\label{tab:events-all}
\end{center}
\end{table}

\begin{table}[p]
\begin{center}
\renewcommand{\arraystretch}{1.2}
\begin{tabular}{|c||r||r|r|r||r|}
\hline
$\sqrt{s}$ & Initial & \multicolumn{3}{|c||}{Overlap with W-Pairs} & Final  \\
 $[\GeV]$  & Sample  & ~$\QQEN$~ & ~$\QQMN$~ & ~$\QQTN$~           & Sample \\
\hline
\hline
$183$      &      86 &        2~ &        9~ &       27~           &  48 \\
$172$      &      15 &        1~ &      ---~ &        3~           &  11 \\
$161$      &       7 &      ---~ &      ---~ &      ---~           &   7 \\
\hline
\end{tabular}
\caption[]{Number of events selected by the hadronic single-W
  selection and the overlap with the $\QQLN$ W-pair selections.
  Duplicate events are removed from the hadronic single-W samples. }
\label{tab:events-sub}
\end{center}
\end{table}

\begin{table}[p]
\begin{center}
\renewcommand{\arraystretch}{1.3}
\begin{tabular}{|l|c||r|r|r|r|}
\hline
Process & $Q^2$ & \giZ & \kg & \Lg & \gvZ \\ 
\hline\hline
$\EEWW$                 & 
$s$                     & 
$+1.13\amp{0.18}{0.18}$ & 
$+1.00\amp{0.39}{0.93}$ &  
$+0.10\amp{0.20}{0.22}$ & 
$-0.44\amp{0.22}{0.23}$ \\
                        &             
                        &             
\small{$(\pm0.13)$}     &
\small{$(\pm0.27)$}     &
\small{$(\pm0.14)$}     &
\small{$(\pm0.17)$}     \\
\hline
$\EE\rightarrow\PW\EN$  &
$\MW^2$                 & 
$+0.57\amp{0.40}{0.93}$ & 
$+1.12\amp{0.31}{0.27}$ &  
$-0.52\amp{0.36}{1.16}$ & 
$-0.55\amp{0.86}{2.24}$ \\
                        &
                        &
\small{$(\pm0.65)$}     &
\small{$(\pm0.34)$}     &
\small{$(\pm0.54)$}     &
\small{$(\pm1.49)$}     \\
\hline        
$\EENN\gamma$           &
$0$                     &
      ---               &
$+1.26\amp{0.96}{0.96}$ &
$+0.41\amp{1.25}{1.26}$ &
      ---               \\
                        &
                        &
   ---                  &
\small{$(\pm1.19)$}     &
\small{$(\pm1.49)$}     &
   ---                  \\
\hline\hline
\multicolumn{2}{|l||}{Combined} &
$+1.11\amp{0.18}{0.19}$ & 
$+1.11\amp{0.25}{0.25}$ & 
$+0.10\amp{0.20}{0.22}$ & 
$-0.44\amp{0.22}{0.23}$ \\
\multicolumn{2}{|l||}{ } &
\small{$(\pm0.13)$}     &
\small{$(\pm0.21)$}     &
\small{$(\pm0.13)$}     &
\small{$(\pm0.17)$}     \\
\hline
\end{tabular}
\caption{Results of one-parameter fits to the TGCs $\giZ$, $\kg$, $\Lg$,
  $\gvZ$, derived from W-pair, single-W and single-photon events, and
  their combination. For each TGC, the other three are set to their SM
  values and the constraints $\DkZ = \Delta g_1^{\mathrm{Z}} - \Dkg
  \tan^2 \theta_w$ and $\LZ = \Lg$ are imposed.  The errors are
  statistical.  Expected statistical errors are given in parenthesis.}
\label{tab:ac-res-1}
\end{center}
\end{table}

\begin{table}[p]
\begin{center}
\renewcommand{\arraystretch}{1.3}
\begin{tabular}{|c||r|r|r|r|r|}
\hline
Parameter  & \giZ   &  \kg &  \Lg  & \kZ & \LZ \\
\hline
\hline
\multicolumn{6}{|c|}{Two-Parameter Fits} \\
\hline
(\giZ,\kg) & $+1.11\amp{0.20}{0.18}$ & $+1.07\amp{0.27}{0.29}$ & --- & --- & --- \\
Corr(\giZ) & $1.00$                  & $-0.24$                 & --- & --- & --- \\
\hline
(\giZ,\Lg) & $+1.18\amp{0.43}{0.23}$ & --- & $-0.08\amp{0.24}{0.48}$ & --- & --- \\
Corr(\giZ) & $1.00$                  & --- & $-0.78$                 & --- & --- \\
\hline
(\kg,\Lg)  & --- & $+1.02\amp{0.30}{0.30}$ & $+0.09\amp{0.21}{0.23}$ & --- & --- \\
Corr(\kg)  & --- & $1.00$                  & $-0.35$                 & --- & --- \\
\hline
\hline
\multicolumn{6}{|c|}{Three-Parameter Fits} \\
\hline
(\giZ,\kg,\Lg)  & 
$+0.97\amp{0.30}{0.35}$ & $+1.07\amp{0.27}{0.26}$ & $+0.13\amp{0.39}{0.28}$ & --- & --- \\
Corr(\giZ)& $ 1.00$  & $-0.21$ & $-0.80$  & --- & ---  \\
Corr(\kg) & $-0.21$  & $ 1.00$ & $ 0.04$  & --- & ---  \\
Corr(\Lg) & $-0.80$  & $ 0.04$ & $ 1.00$  & --- & ---  \\
\hline
(\giZ,\kg,\kZ)  &
$+1.80\amp{1.23}{0.45}$ & $+1.07\amp{0.24}{0.23}$ & --- & $+0.41\amp{0.53}{1.16}$ & --- \\
Corr(\giZ)& $ 1.00$ & $ 0.07$ & --- & $-0.57$ & --- \\
Corr(\kg) & $ 0.07$ & $ 1.00$ & --- & $-0.29$ & --- \\
Corr(\kZ) & $-0.57$ & $-0.29$ & --- & $ 1.00$ & --- \\
\hline
\hline
\multicolumn{6}{|c|}{Four-Parameter Fit} \\
\hline
(\kg,\Lg,\kZ,\LZ) & --- & 
$+1.20\amp{0.26}{0.37}$ & $+0.42\amp{0.57}{0.34}$ & $+1.23\amp{0.42}{0.40}$ & $-0.46\amp{0.34}{0.26}$ \\
Corr(\kg) & --- & $ 1.00$ & $-0.35$ & $-0.29$ & $-0.30$ \\
Corr(\Lg) & --- & $-0.35$ & $ 1.00$ & $-0.10$ & $-0.05$ \\
Corr(\kZ) & --- & $-0.29$ & $-0.10$ & $ 1.00$ & $-0.26$ \\
Corr(\LZ) & --- & $-0.30$ & $-0.05$ & $-0.26$ & $ 1.00$ \\
\hline
\end{tabular}
\caption[]{
  Results on the C- and P-conserving TGCs derived from the two- and
  three-parameter fits to $(\giZ,\kg)$, $(\giZ,\Lg)$, $(\kg,\Lg)$, and
  $(\giZ,\kg,\Lg)$, imposing the constraints $\gvZ=0$, $\DkZ = \Delta
  g_1^{\mathrm{Z}} - \Dkg \tan^2 \theta_w$ and $\LZ = \Lg$; from the
  three-parameter fit to $(\giZ,\kg,\kZ)$ imposing the constraints
  $\gvZ=\Lg=\LZ=0$; and from the four-parameter fit to
  $(\kg,\Lg,\kZ,\LZ)$ imposing the constraints $\gvZ=0$ and $\giZ=1$.
  The matrices of correlation coefficients are also given.  The errors
  are statistical, combining all processes. }
\label{tab:ac-res-2}
\end{center}
\end{table}

\begin{table}[p]
\begin{center}
\renewcommand{\arraystretch}{1.2}
\begin{tabular}{|l||r|r|r|r|}
\hline
Process     & $\giZ$ & $\kg$ & $\Lg$ & $\gvZ$ \\
\hline    
\hline    
$\PW\PW$    &   0.10 &  0.39 &  0.08 &   0.12 \\
$\PW\EN$    &   0.33 &  0.26 &  0.32 &   0.86 \\
$\NN\gamma$ &    --- &  0.90 &  0.99 &    --- \\
\hline
Combined    &   0.10 &  0.17 &  0.10 &   0.12 \\
\hline
\end{tabular}
\caption[]{ Systematic errors in the determination of
  the TGCs $\giZ$, $\kg$, $\Lg$ and $\gvZ$ for the individual
  processes and their combination.  For each TGC, the other three are
  set to their SM values and the constraints $\DkZ = \Delta
  g_1^{\mathrm{Z}} - \Dkg \tan^2 \theta_w$ and $\LZ = \Lg$ are
  imposed. }
\label{tab:ac-syst-1}
\end{center}
\end{table}

\begin{table}[p]
\begin{center}
\renewcommand{\arraystretch}{1.2}
\begin{tabular}{|l||r|r|r|r|r|}
\hline
Process     & $\giZ$ & $\kg$ & $\Lg$ & $\kZ$  & $\LZ$ \\
\hline    
\hline    
$\PW\PW$    &   0.09 &  0.46 &  0.10 &   0.25 &  0.10 \\
$\PW\EN$    &   0.35 &  0.32 &  0.50 &   0.50 &  0.63 \\
$\NN\gamma$ &    --- &  0.90 &  0.99 &    --- &   --- \\
\hline
Combined    &   0.10 &  0.28 &  0.10 &   0.23 &  0.11 \\
\hline
\end{tabular}
\caption[]{Systematic errors in the determination of
  the TGCs $\giZ$, $\kg$, $\Lg$, $\kZ$ and $\LZ$ for the individual
  processes and their combination.  For each TGC, all other TGCs,
  including $\gvZ$, are set to their SM values. }
\label{tab:ac-syst-2}
\end{center}
\end{table}

\end{document}